\documentclass[preprint,12pt]{elsarticle}

\usepackage{amssymb}
\usepackage{xcolor}
\usepackage{amsmath}
\usepackage{url}
\usepackage{mdframed}
\usepackage{booktabs} 

\usepackage{xcolor}
\usepackage{graphicx}
\usepackage{subcaption}
\usepackage{tikz}
\usetikzlibrary{fit,calc}
\usepackage{mdframed}

\journal{Elsevier Journal}

\begin{document}

\begin{frontmatter}



\title{Dynamics of a Tuberculosis Outbreak Model in a Multi-scale Environment}


\author[inst1,inst2]{Selain K. Kasereka}

\affiliation[inst1]{organization={University of Kinshasa, Faculty of Sciences and Technologies, Department of Mathematics, Statistics and Computer Science},
            postcode={Kinshasa XI P.O. Box 190},
            country={Democratic Republic of the Congo}}
\affiliation[inst2]{organization={ABIL Research Center},
            postcode={Kinshasa XI P.O. Box 190},
            country={Democratic Republic of the Congo}}

\begin{abstract}
Modeling and simulation approaches for infectious disease dynamics have proven to be essential tools for effective control of the spread of epidemics in the population. Among these approaches, it is obvious that compartmental mathematical models, such as SIS, SIR, SEIR, etc. are the most widely used by researchers. However, they are difficult to apply in a multi-scale environment, especially if we want to take into account the heterogeneous behaviors of individuals. The aim of this paper is to present a hybrid model in which an Equation-Based Model (EBM) of tuberculosis dynamics is coupled to an Agent-Based Model (ABM) in a two-scale environment. In this model, individuals are placed in cities considered as agents in which the dynamics of the disease is modeled by eight compartments and managed by solving a system of differential equations. Individual agents move between these cities using an ABM that controls their mobility. Considering some parametric values and assumptions, the results obtained show that human mobility has a significant impact on the spread of tuberculosis within the population.  The management of population and disease dynamics at different levels (microscopic and macroscopic) testifies to the robustness of the proposed approach.
\end{abstract}




\end{frontmatter}


\section{Introduction}

Despite being both preventable and curable, tuberculosis (TB) remains one of the world’s leading infectious disease killers according to the World Health Organization's (WHO) report~\cite{world2022global}. An estimated 10.6 million people fell ill with TB in 2021, up from 10.1 million in 2020, and 1.6 million people died of TB in 2021. The incidence rate of tuberculosis increased by 3.6\% in 2021 compared to 2020~\cite{bagcchi2023s}. This disease is spread mainly by air when infected people cough or sneeze. It is caused by the bacterium Mycobacterium tuberculosis~\cite{world2022global}. Poverty and rural exodus are key factors in the spread of tuberculosis. Rural exodus, with migration to urban areas, often leads to overcrowded neighborhoods with limited access to health services, and rural migrants often work in precarious conditions without social protection, increasing their risk of exposure to tuberculosis. Limited access to healthcare, lack of diagnosis and treatment, and stigmatization are major obstacles in the fight against tuberculosis.

The increase in the number of cases reported worldwide as reported in \cite{bagcchi2023s}, calls for an enhanced response in the fight against tuberculosis, which requires in-depth scientific research. It is therefore imperative to develop mechanisms for understanding the spread of this disease within the population at different scales, using modeling and simulation approaches that take into account population mobility. 

In recent years, compartmental mathematical models such as SIS, SIR, SEIR, etc. have been the most widely used to control and understand the dynamics of infectious diseases in populations ~\cite{atangana2014computational, cao2023dynamics, doungmo2014some, goufo2016stability, kasereka2014hybrid, kasereka2018estimation, ojo2023mathematical, trauer2014construction,  zhao2017analysis}. However, the concrete application of these models to environments that require the consideration of individual mobility is not easy. That's why meta-population models have emerged, they consider that populations of the same species are spatially distributed and that there are more or less regular and important exchanges of individuals between these populations~\cite{hanski2004metapopulation}. The habitat of this metapopulation is an ecological unit corresponding to the landscape, i.e. a set of sites with various stages of ecological succession and whose geography allows limited gene exchanges but existing from one site to another. In general, metapopulation processes emerge within a population of individuals as a result of the spatial fragmentation of their habitat. These models have increased their weakness with regard to the mobility components. These limitations are treated in an aggregate manner, and therefore do not consider the heterogeneous behaviors of individuals.  

In the literature, several researchers have been interested in agent-based models (ABM) derived from artificial intelligence to understand diseases transmission in the population~\cite{chen2024role,chumachenko2024exploring,gallagher2024epidemiological,modu2023agent}. With regard to the spread of tuberculosis, these models have been applied in various research~\cite{kabunga2020stochastic,petrucciani2024silico,udall2023modelling}. These models have made it possible to model interactions and contacts between individuals (infected or not) in order to take into account certain complexities of social systems that deterministic models are unable to capture. Drawing on the advantages of equation-based models on the one hand, and agent-based models on the other, a number of researchers have demonstrated the need to couple these two approaches~\cite{avegliano2023equation,kasereka2014hybrid,minucci2024agent}. These studies have shown that coupling these approaches should facilitate understanding of the dynamics of complex systems, and enable efficient calibration of parameters at different levels. 

This paper proposes a hybrid model in which a compartmental mathematical model of tuberculosis is coupled to an agent-based model to analyze the spread of the disease at different scales. Indeed, individuals are placed in cities in which the dynamics of the disease is managed by solving the differential equation system by the Runge-Kutta method and individuals move between several cities over a circle using an agent-based model. Simulations of the hybrid model proposed are performed and the impact of population mobility on the spread of the disease is assessed. 

The remainder of this paper is organized as follows: We present the description of the mathematical model (See Section~\ref{Math_description}), then we show the description of the ABM using the Overview, Design concepts and Details (ODD) protocol  (See Section~\ref{Agent_description}). Section \ref{Simulation} presents performed simulations before discussing the results obtained in Section~\ref{Discussion}. Finally Section~\ref{Conclusion} is devoted to the concluding remarks.

\section{Description of the Mathematical Model \label{Math_description}}

We consider a compartmental model with 8 compartments (groups). In the model we consider a population $S$ that is susceptible to contract TB infection. This population can be infected according to a contact rate $\alpha$ and a transmission rate $\lambda$. For this, there is a proportion $1-p$ of this population that will be infected and therefore will be part of compartment $I$, this is a fast progression to the active TB. We note that a susceptible individual can become latent $L_e$ (latent early) following a contact rate $\alpha$ and a transmission rate $\lambda$. In this model a latent individual ($L_e$ and $L_f$) is not yet able to transmit the disease. Latent individual $L_e$ can become $L_f$ (latent late) following a given rate $h$. A latent $L_e$ can also directly become infectious (able to infect other people) at a rate $q$, an individual $L_f$ can also become infectious $I$ at a rate $w$, this is the low progression to the active TB. 

In this model, infected and infectious individuals, who are in the $I$, $L_e$ and $L_f$ compartments can heal spontaneously and move in the compartment $R_2$ according respectively to the rates $\sigma$, $g_2$ and $k_2$. They can also heal after treatment process and move in the compartment $R_1$ according respectively to the rates $\gamma$, $g_1$ and $k_1$. These healed individuals can be re-infected according to a rate of transmission $\lambda$, a contact rate $\alpha$ and a re-infection rate $r$. Infectious people $I$, under treatment can also become $L_e$ according to a rate $r_1$. During the treatment process, there are people who are lost to follow up, so people who stop treatment ($K$). These individuals can be re-infected according to a rate $r_3$. In the model, we consider other people who are transferred to other hospitals for lack of capacity or medication ($T$). These people can be re-infected at a rate $r_2$. 

Demography is considered in the TB proposed model. We note $\Lambda$ the rate of recruitment of susceptible individuals $S$. In the model people can die, for that we consider $\mu_1$ as the rate of natural mortality (not related to TB infection) and $\mu_2$ the rate of mortality linked to TB infection. The Figure~\ref{modelTB} below presents the TB transmission dynamics between the different compartments of the model.

\begin{figure}[h!]
\label{modeltba}
\begin{center}
\includegraphics[width=120mm, height=65mm]{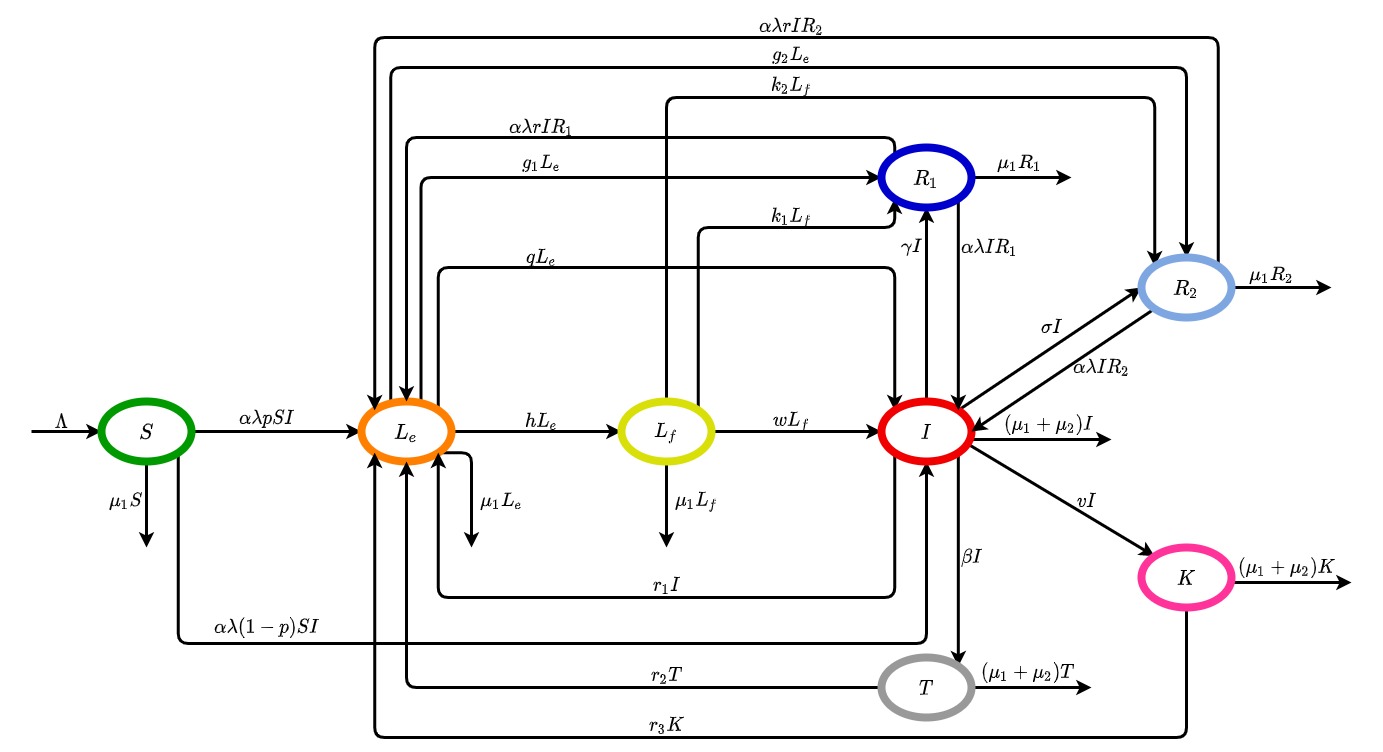}
\caption {Compartmental model of TB transmission.}
\end{center}
\end{figure}

Based on presented information, we obtain the Ordinary Differential Equation System Eq. \eqref{modelTB}.

\begin{equation} \left\{
\label{modelTB}
\begin{array}{r@{~}l@{~}l}
\dot{S}&=&\Lambda-\alpha \lambda p SI-\alpha \lambda (1-p)SI-\mu_1 S \\
\dot{L_e}&=&\alpha \lambda p S I+ \alpha \lambda r I(R_1+R_2)+r_1 I + r_2 T + r_3 K - \widetilde{B} L_e \\ 
\dot{L_f}&=&hL_e-(\mu_1+w+ k_1 + k_2)L_f \\
\dot{I}&=& w L_f+qL_e- \widetilde{A} I+\alpha \lambda R_1 I +\alpha \lambda R_2 I +\alpha \lambda(1-p)SI \\ 
\dot{R_1}&=& g_1 L_e + k_1 L_f + \gamma I- \alpha \lambda r R_1 I - \mu_1 R_1 - \alpha \lambda R_1 I \\
\dot{R_2}&=&\sigma I + k_2 L_f + g_2 L_e - \alpha \lambda R_2 I - \alpha \lambda r R_2 I - \mu_1 R_2 \\
\dot{T}&=&\beta I-(\mu_1+\mu_2 + r_2)T \\ 
\dot{K}&=&v I-(\mu_1+\mu_2 + r_3)K \\ 
N&=&S+L_e+L_f+I+R_1+R_2+T+K
\end{array}
\right.\end{equation}
Where $\widetilde{A}= (r_1 +\gamma+\beta+\sigma+ v +\mu_1+\mu_2)$ and $\widetilde{B}=(\mu_1+h+q+g_1 + g_2)$.\\

The positivity of the solution, the existence and uniqueness of the solution,  the calculation of equilibrium points (DFE and EE), the basic reproduction number $\mathcal R_0$ and the stability of the equilibrium points of this model is presented in our recent work~\cite{kasereka2020analysis}.

The dynamics of the disease at the level of all cities is managed by solving the system of differential equations to control the dynamics of the disease. This is the macroscopic level of the model. For individuals to move from city to city, they are managed by a model based on intelligent agents that manages each agent's behaviors, including the probability of deciding to travel and contamination during the trip. Once the individual arrives at his destination, he contributes to the epidemiology of the disease in the city. This is the microscopic level of the model. Note that at each simulation step, cities generate individuals with their status according  to a fixed rate (see Table~\ref{Tab:parameters}).   

\section{Description of the ABM using ODD Protocol \label{Agent_description}}

To describe the agent-based model, we used the ODD (Overview, Design concepts, and Details) protocol as a coherent, logical and readable account of ABM structures and dynamics~\cite{grimm2020odd}.

\subsection{Overview}

\begin{itemize}
	\item \textit{Purpose}: the purpose of the model was to assess the impact of population mobility on the spread of TB in a multi-scale environment.
	
	\item  \textit{Entities, state variables and scale}: (1) \textit{Agent}: in this model we consider two types of agent. The first one is the individual. The second one is the city. Indeed, an individual agent is located in a given agent city. Each individual agent has state variables as well as characteristics presented in Table~\ref{Tab:stateVhybrid1}. (2) \textit{Spatial Units}: agent cities generate individual agents according to a given rate of generation. An individual agents can move from a city to others. State variables of cities are listed in Table~\ref{Tab:stateVhybrid2}. 
	
	\item \textit{Environment}: we assume that cities have the same population size. Based on the geometry of the environment in Gama Platform~\cite{grignard2013gama}, we set the size of a city at $50 \times 50$. The location of cities is a circle and is calculated by the function presented by Eq. \eqref{fonctionEnv}.
	
	\begin{equation}
	\label{fonctionEnv}
	L(v) = f \left(\frac{c_1}{2} + r_x cos \left(\frac{360}{b} \times i \right), \frac{c_2}{2} + r_x sin \left(\frac{360}{b} \times i \right)\right)  
	\end{equation}
	
	Where

 \begin{equation*}
	r_x = \left(\frac{c_1 - s \times 2}{2}\right)
\end{equation*}

Key Components:

\begin{itemize}
\item $i$: The city number or identifier.
\item $s$: The size of the city.
\item $c_1$: The length of the environment.
\item $c_2$: The width of the environment.
\item $b$: The number of cities.
\item $f$: A location function in a two-dimensional plane.
\end{itemize}
	
	At each simulation time step, each city generates individual agents according to their states (susceptible, early latent, late latent, infectious, spontaneously recovered, recovered after treatment, transferred or lost to follow- up). The generation rate of this population is given by the variable $z$ which is randomly selected between 0.001\% and 1\% closed interval.
	
	\item \textit{Process overview and scheduling}: the model process is in discrete time steps representing 1 hour. At the beginning of the simulation, in cities, individual agents are in any state. The disease dynamics is managed by solving a differential equation system using RK4 method as applied in~\cite{kasereka2014hybrid}. Individual agents have a probability of leaving a city to another. At each time step the sub-models are executed (movement, becoming infected, healing, updating of global variables, etc.). 
	
\end{itemize}

\begin{table}[ht!]
	\footnotesize
	\caption {Description of agent state variables for the hybrid model.}
	\label{Tab:stateVhybrid1}
	\begin{center}
		\begin{tabular}{ll}
			\hline
			\textbf{State variables}	 & \textbf{Description}\\  
			\hline
			Agent identifier &	Unique agent id\\
			Age	& From 0 to 120\\
			Sex	& Male, Female\\
			Susceptible? &	Is the agent susceptible?\\
			Early latent? &	Is the agent early latent?\\
			Late latent? &	Is the agent latent late? \\
			Sick?	&Is the agent sick?\\
			Spontaneously recovered?&	Is the agent spontaneously recovered?\\
			Recovery after treatment?&	Is the agent recovered after treatment?\\
			Transferred?&	Is the agent transferred?\\
			Lost to follow-up?&	Is the agent lost to follow-up?\\
			Sick days &	Number of days the agent has been sick\\
			Early latent days&	Number of days the agent has been early latent\\
			Latent late days&	Number of days the agent has been late latent \\
			Transferred days&	Number of days the agent has been transferred\\
			Lost to follow-up days&	Number of days the agent has been Lost to follow-up\\ 
			Spontaneously recovered days?&	Number of days the agent has been  spontaneously recovered?\\
			Recovery after treatment days?&	Number of days the agent has been recovered after treatment?\\
			Location& 	Determines the location of the agent on the environment\\
			In movement?&	 Is the agent in movement?\\
			Neighbors	& Represents the list of the agent's neighbors\\
			Trip range of contact & Represents the radius of contamination of the agent during\\ &the trip. The default is taken between 1 and 5m.\\
			Trip in contact with sick people?&	Determine if the agent is in contact with sick people in its \\ &range during the trip.\\ 	
			Destination city &	The city the agent is moving towards\\
			Trip contact number	& List of the number of contacts an agent has had during the trip\\
			Trip average contacts &	The average number of contacts an agent has had during the trip\\
			Trip contact type &	Random\\
			\hline
		\end{tabular}
	\end{center}
\end{table}

\begin{table}[ht!]
	\footnotesize
	\caption {Description of city state variables.}
	\label{Tab:stateVhybrid2}
	\begin{center}
		\begin{tabular}{ll}
			\hline
			\textbf{State variables}	 & \textbf{Description}\\  
			\hline
			City identifier	&Unique city id\\
			Total individual agent	&The total number of agents in the city\\
			Time generate agent & The total number of agents generated each time step\\
			\hline
		\end{tabular}
	\end{center}
\end{table}

\subsection{Design Concept}

\begin{itemize}
	\item \textit{Basic principle}: the dynamics of TB is considered at two levels. (1) The first one is in the city (macroscopic). Here, the spread of the disease is based on mathematical model (Ordinary Differential Equation). For that, eight compartments are considered (susceptible, early latent, late latent, infectious, spontaneously recovered, recovered after treatment, transferred and lost to follow-up). In this model, when a susceptible individual comes into contact with an infectious individual, it is possible that the susceptible individual may become exposed or infectious according to a rate of contact and a rate of transmission. People can travel from their cities to others by choosing a random destination. 
	(2)	The second one is the individual (microscopic). Here, once individuals travel from their hometown to another city, they are considered as agents. Indeed, during the trip, the dynamics of the disease is managed by an agent-based model. 
	
	\item \textit{Emergence}: the emergence of the system can be seen in the evolution of TB infection in all cities considered. At the macroscopic level, it depends on the contact rate and transmission rate between infectious people and susceptible people. At the microscopic level, it depends on the type of initially infectious agents, the other agents that come into contact with them along a radius of contamination, the duration of contact but also the frequency of contact. 
	
	\item \textit{Adaptation}: infectious agents can detect susceptible and cured agents. Once cured agents detect infectious agents, these agents avoid contact with them. 
	
	\item \textit{Interaction}:  in this model we assume that agents within the same radius interact with each other and also with their environment. For example, if a susceptible agent is in a same radius as an infectious agent, it is possible that the susceptible agent become contaminated. 
	
	\item \textit{Stochasticity}: The movement of the agents with their specific states are stochastic. The choice of the destination city is random. Stochasticity is also in the TB contamination as applied in~\cite{kabunga2020stochastic}. During the traveling, if a susceptible agent comes into contact with an infectious agent, there is a certain probability that determines whether that agent becomes exposed or directly infectious. In addition, the length of time an agent remains in a state (susceptible, early latent, latent late, infectious, spontaneously recovered, recovered after treatment, transferred and lost to follow-up) is randomly chosen.
	
	The probability of leaving one city $A$ to another city $B$ is calculated by Eq. \eqref{Eq:ProbaTravel}.

\begin{equation}
P(travel)=P(h_{AB})
\label{Eq:ProbaTravel}
\end{equation}
 
With $A$ the source city and $B$ the destination city, $h$ a value between $0.1$ and $0.9$. 

In this model, individual leaving his or her home city to a randomly chosen destination city must pass through a central point of the circle called "hub". The mobility model of individuals on the transport network is managed by an agent-based model.
	
	\item \textit{Observation}: data are collected at each execution of the model on individual agents according to their states (susceptible, early latent, latent late, infectious, spontaneously recovered, recovered after treatment, transferred and lost to follow-up). The output of the model is collected at each time step to observe the evolution of infectious agents over time.
\end{itemize}

\subsection{Details }

\begin{itemize}
	\item \textit{Initialization}: at the beginning of the simulation, a number of susceptible individuals is placed in all cities and only one city has in addition some infectious individuals. 
	
	\item \textit{Sub-model:}
	
	\begin{itemize}
		\item \textit{Clock}: this sub-model manages time in the model. It is determined in hour, day, week, month and year. After every 24 hours, the hour variable is reset to zero and the day variable is incremented by 1. After every 7 days, the day variable is reset to zero and the week variable is incremented by 1. After every 4 weeks, the week variable is reset to zero and the month variable is incremented by 1. After every 12 months, the month variable is reset to zero and the year variable is incremented by 1. We therefore assume a month to 4 weeks (28 days). 
		
		\item \textit{Mobility}: agents move from a city to another. The choice of destination city is random. All information presented below is considered during the trip of the agent from a city to another.
		
		\item \textit{Contamination}: at the microscopic level, contamination is based on contact between a susceptible agent and an infectious agent, and the probability of TB transmission. If a susceptible agent is infected then he will change his status from susceptible to exposed (Early latent ($L_e$) / Late latent  ($L_f$)) or infectious. At the macroscopic level, contamination is managed by the resolution of the differential equation using the fourth-order Runge-Kutta method (RK4).
		
		\item \textit{Early latent}: An exposed agent ($L_e$) progresses to ($L_f$) or to infectious according to a certain probability of progression. The time that the agent will remain in the state $L_e$ before moving to $L_f$ is determined by a certain probability as well.
		
		\item \textit{Late latent }: an exposed agent ($L_f$) progresses to infectious according to a certain probability of progression. The length of time the agent will remain in the ($L_f$) state before becoming infectious is also determined by a certain probability.
		
		\item \textit{Recover:} infectious agents recover based on a probability of spontaneous recovery or a probability of recovery after taking medication. 
		
		\item \textit{Relapse:} cured agents may relapse and become exposed or infectious based on a certain probability. 
		
		\item \textit{Lost to follow-up:} when an infectious agent stops taking medication before the end of treatment, it is considered lost to follow-up. Infectious agents become lost to follow-up with a certain probability.
		
		\item \textit{Global variables update:} all global variables are updated at the end of each time step. At the same time also the number and percentage of susceptible, exposed (Early latent and Latent late), infected, transferred, lost to follow-up, Recovered (spontaneously and after treatment) agents are all calculated. 
	\end{itemize}
\end{itemize}

\section{Simulations of the Hybrid Model \label{Simulation}}

To implement the resulting hybrid model, we used a special language called GAML (GAma Modeling Language) provided by the GAMA platform, which is an open source modeling and simulation environment for creating spatially explicit agent-based simulations~\cite{grignard2013gama}.

\subsection{Parameters used in the Model}

The Table~\ref{Tab:parameters} presents used parameters in performed simulations. Most of them was taken from the literature. Other parameters was assumed based on data at our disposal.

\begin{table}[ht!]
\footnotesize
\caption {Parameter values and their meanings}
\label{Tab:parameters}
\begin{center}
\begin{tabular}{lllll}
\hline
Prms  & Meaning  & Value  & Reference	& Assumed     \\  
\hline
$\Lambda$ & Rate of recruitment ($\Lambda \times N$)  & 0.0100 &  &  Yes  \\
$\mu_1$ & Natural death rate& 0.0222 &~\cite{ozcaglar2012epidemiological} &  \\
$\mu_2$ & Mortality rate linked to TB & 0.040 &~\cite{bisuta2018tendances} &  \\
$\gamma$ & Recovered rate after treatment (I to $R_1$) & 0.840 & ~\cite{bisuta2018tendances} & \\
$\sigma$ & Spontaneously recovered rate (I to $R_2$) & 0.250 &  ~\cite{kasereka2020analysis}& \\
$\alpha$ & Contact rate & 0.0010  &  & Yes \\
$\lambda$ & Rate of transmission & 0.100 &~\cite{kasereka2020analysis} &  \\
$1-p$ & Fraction of fast-developing active TB& 0.05 &~\cite{blower1995intrinsic}~\cite{zhao2017analysis}& \\
$\beta$ & Rate of transfer  to a hospital & 0.010 &  & Yes\\
$v$ &  Rate of lost to follow up & 0.030 &  ~\cite{bisuta2018tendances}& \\
$q$ & Progression rate ($L_e$ to $I$)& 0.129 & ~\cite{trauer2014construction} &\\
$h$ & Rate of progression of TB ($L_e$ to $L_f$) & 0.821 & ~\cite{trauer2014construction} & \\
$r$ & Reinfection rate ($R_i$ to $L_e$) with i=1,2& 0.030 & ~\cite{blower1995intrinsic} & \\
$r_1$ & Rate of re-infection (I to $L_e$) & 0.63 & ~\cite{trauer2014construction} & \\
$r_2$ & Rate of re-infection (T to $L_e$) & 0.63 & ~\cite{trauer2014construction} & \\
$r_3$ & Rate of re-infection (K to $L_e$) & 0.63 & ~\cite{trauer2014construction} & \\
$g_1$ & Rate of recovered ($L_e$ to $R_1$) & 0.840 &  ~\cite{bisuta2018tendances} &\\
$g_2$ & Rate of spontaneously recovered ($L_e$ to $R_2$) & 0.250 & ~\cite{kasereka2020analysis} &\\
$k_1$ & Rate of recovered ($L_f$ to $R_1$) & 0.840 &  ~\cite{bisuta2018tendances}  &\\
$k_2$ & Rate of spontaneously recovered ($L_f$ to $R_2$) & 0.250 & ~\cite{kasereka2020analysis} & \\
$w$ & Rate of progression ($L_f$ to I) & 0.075 & ~\cite{trauer2014construction} & \\
$m$ & Mobility rate  & 0.00001 - 0.01 & &  Yes\\
$gen^\ast$ & Individual generation rate in city  & 0.00001 - 0.01 & &  Yes\\
\hline
\end{tabular}
\end{center}
\end{table}

\subsection{Numerical Simulation}

To simulate this model we will consider the following case: First, we consider 6 cities and a high mobility of people between those cities. Second, we simulate the model by varying the mobility rate of individuals (low mobility rate for infected and infectious individuals) between 6 cities. 

This model is implemented using the GAMA platform which is a platform designed for field experts, modelers and computer scientists. It is therefore a complete modeling and simulation development environment for the construction of spatially explicit multi-agent simulations. This platform was designed completely in Java. As a free tool, GAMA uses the GAML (GAma Modeling Language) for coding~\cite{grignard2013gama}.

For all simulations, we consider that the disease exists in one of the cities at the beginning of the simulation. This allows us to evaluate the impact of population mobility on the disease dynamics (TB).

\subsubsection{\textit{Simulation with a High Mobility Rate for all Individuals}}

\begin{figure}
	\centering
	\begin{tabular}{cccc} 
		\includegraphics[scale=0.2]{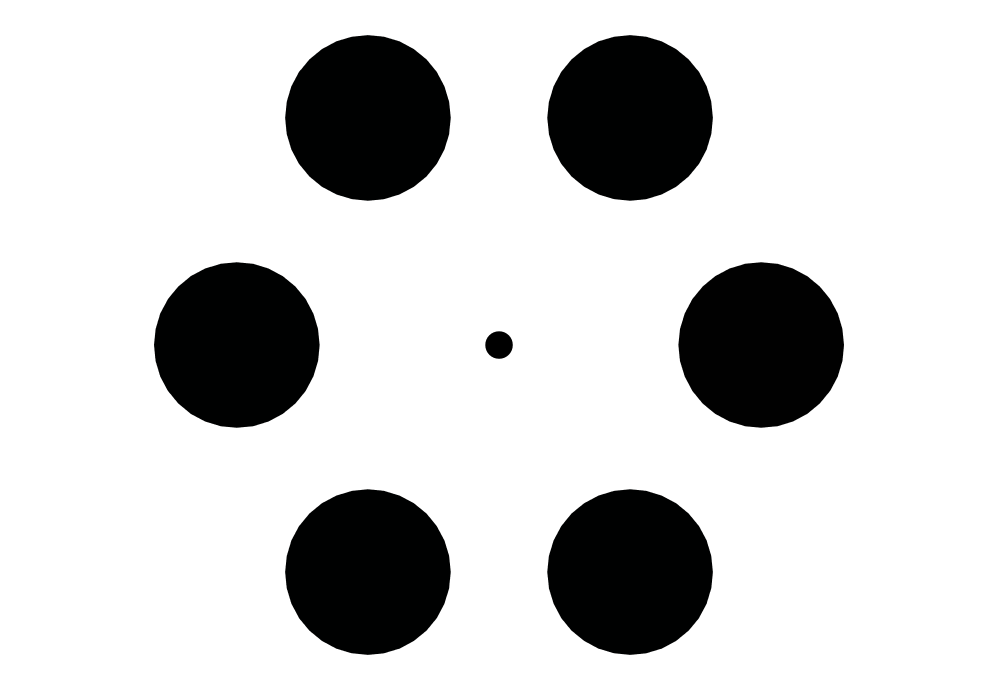}&
		\includegraphics[scale=0.2]{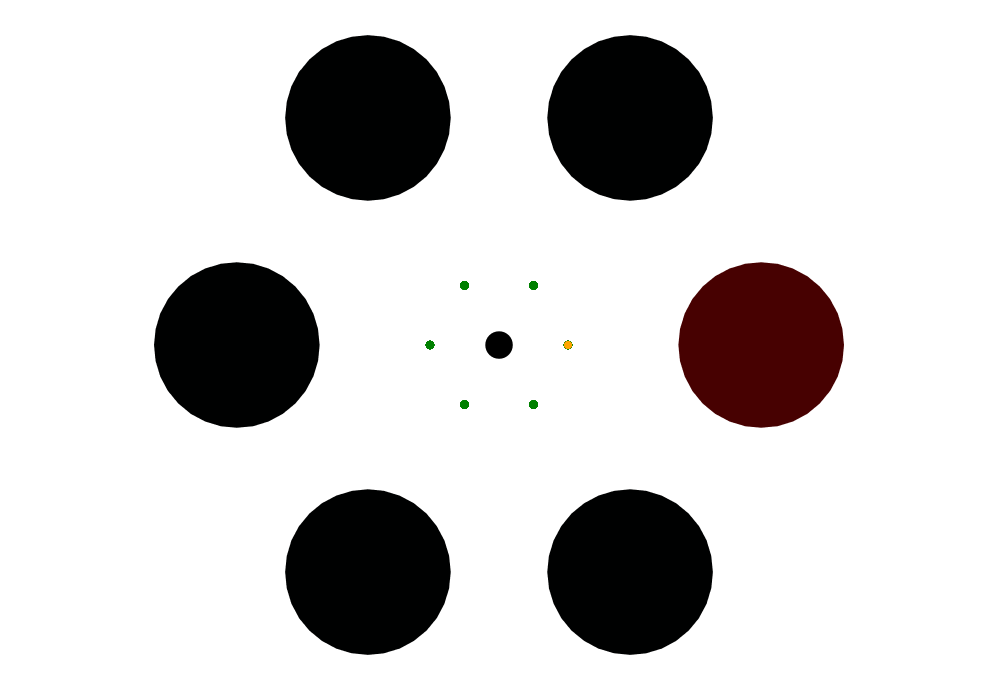} \\
		(a) & (b) \\
		\includegraphics[scale=0.2]{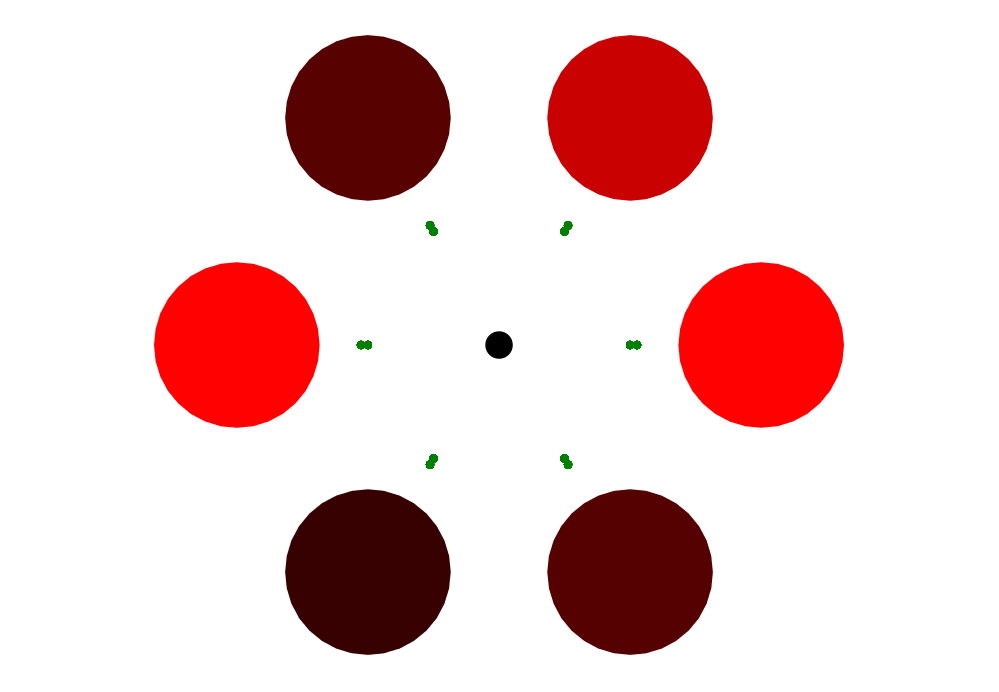}&
		\includegraphics[scale=0.2]{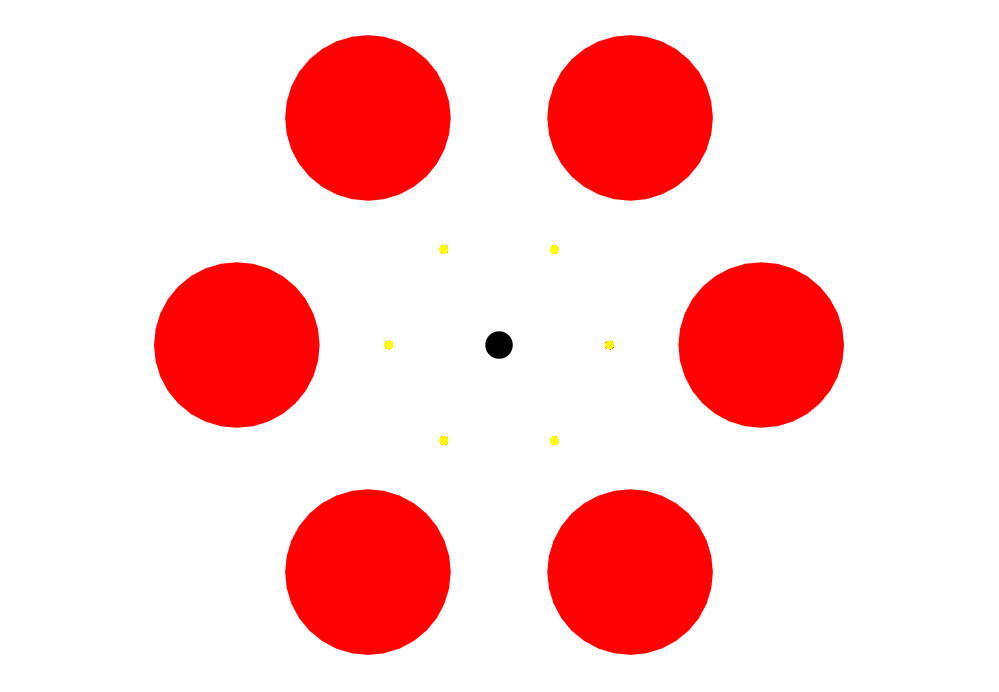} \\
		(c) & (d) \\
	\end{tabular}
	\caption{Presentation of the environment with 6 cities (circles) interconnected via a hub. (a) simulation at time $t=0$, (b) simulation at time $t=79$, (c) simulation at time $t=262$, (d) simulation at time $t=328$. The concentration of the red color of the city is proportional to the level of infectivity of the city. \label{fig:villes1}}
\end{figure}

By simulating the proposed model with the basic parameters shown in Table~\ref{Tab:parameters}. We consider the dynamics of TB which spreads globally over 6 connected cities with the mobility rate of all individual ($S, L_e, L_f, I, T, K, R_1$ and $R_2$) equal to $0.01$. Figure~\ref{fig:villes1} shows the evolution of the model during simulation. 

\begin{figure}[ht!]
	\centering
	\begin{tabular}{cccccc} 
		\includegraphics[scale=0.35]{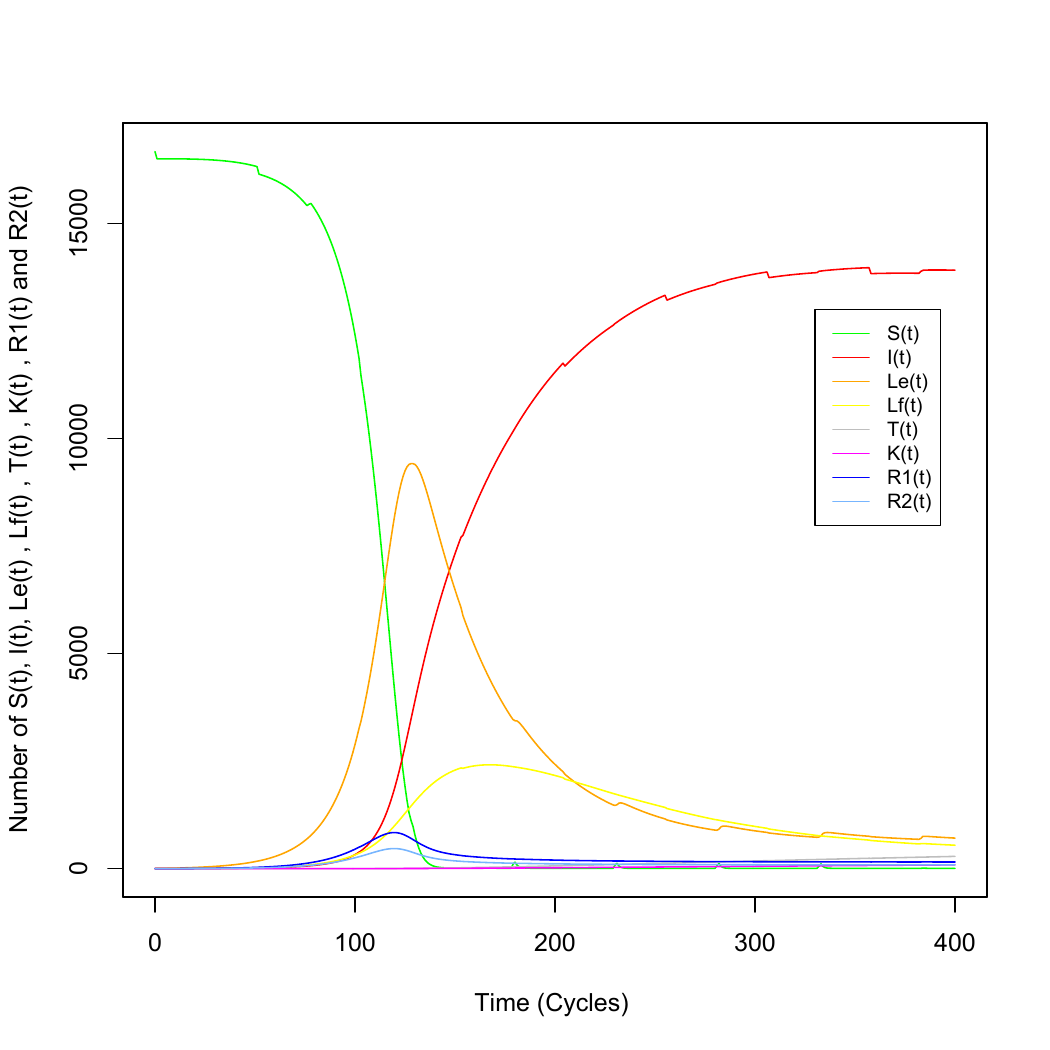}&
		\includegraphics[scale=0.35]{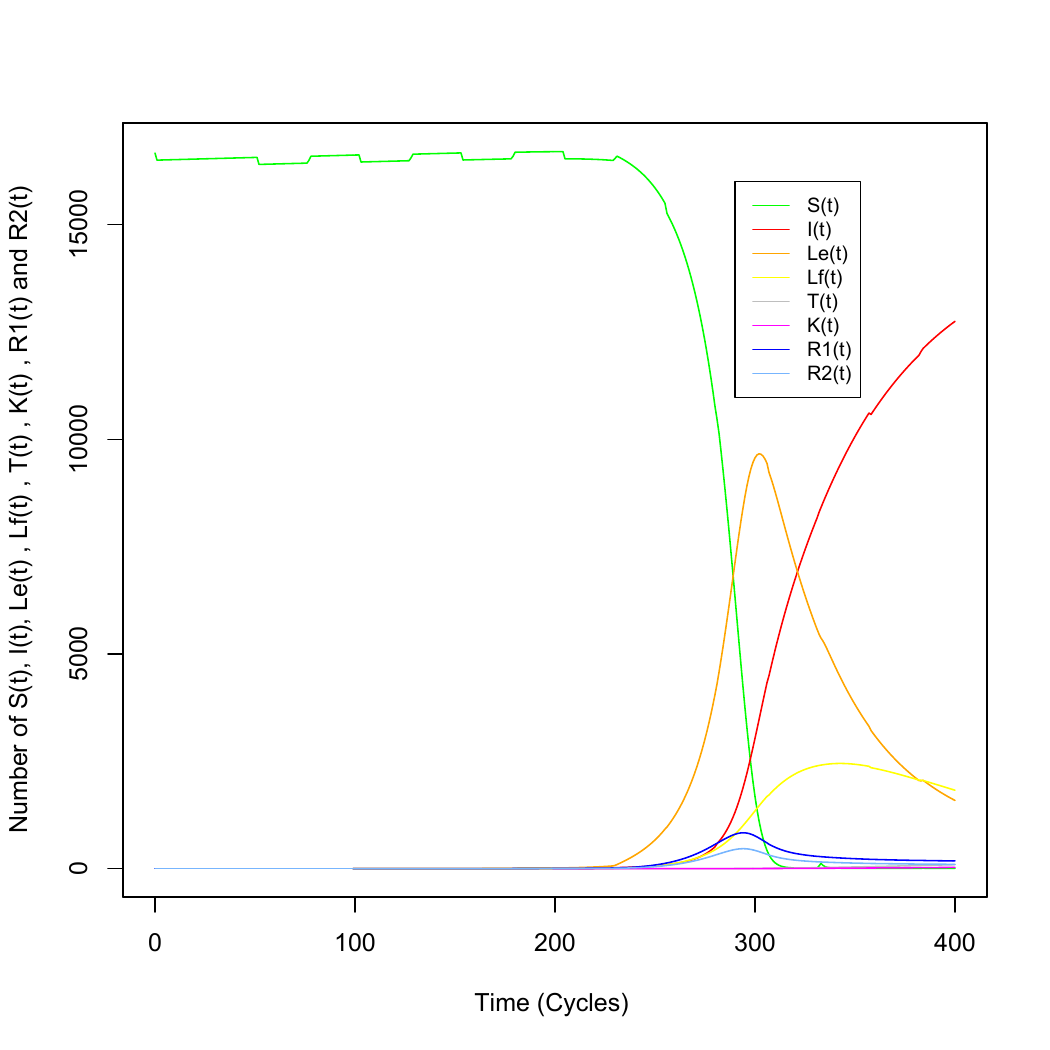} \\
		(a) & (b) \\
		\includegraphics[scale=0.35]{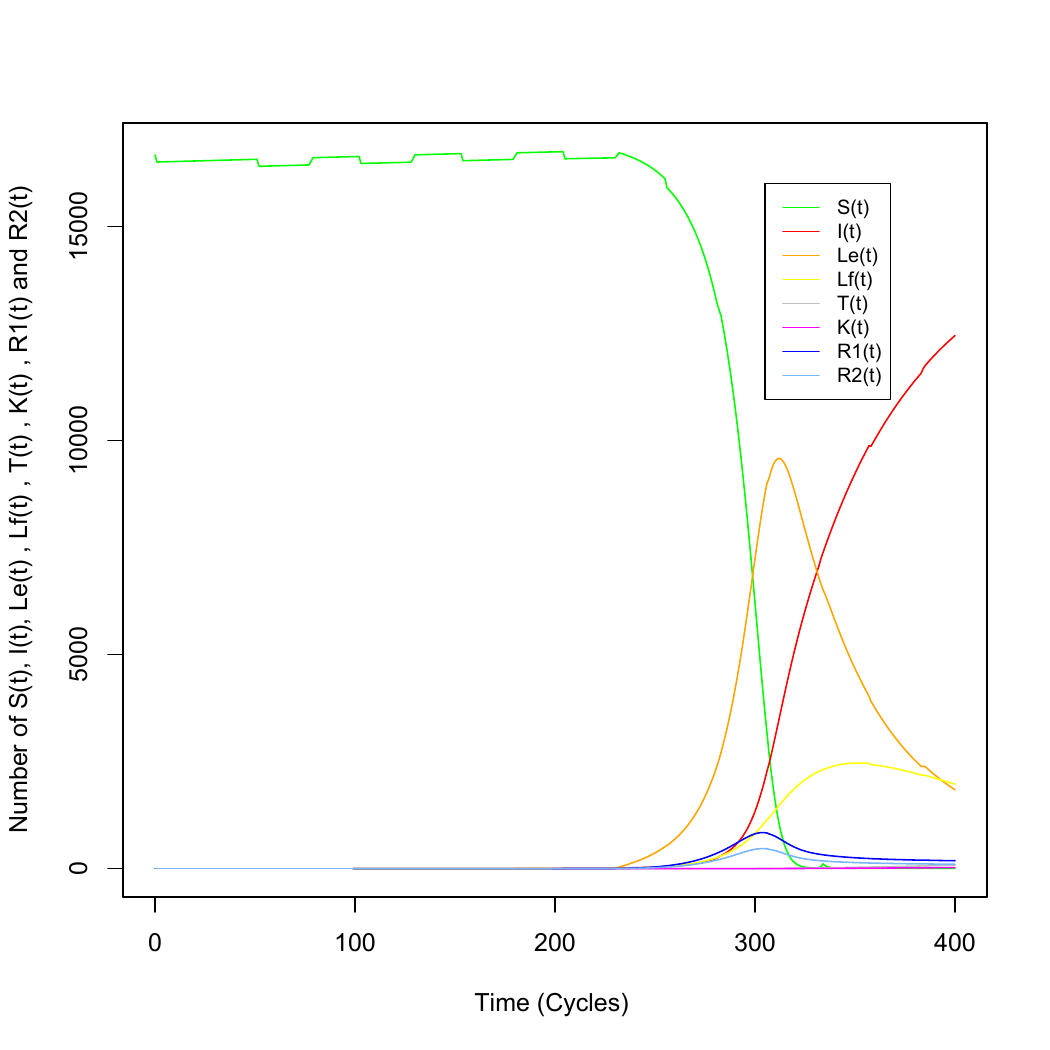}&
		\includegraphics[scale=0.35]{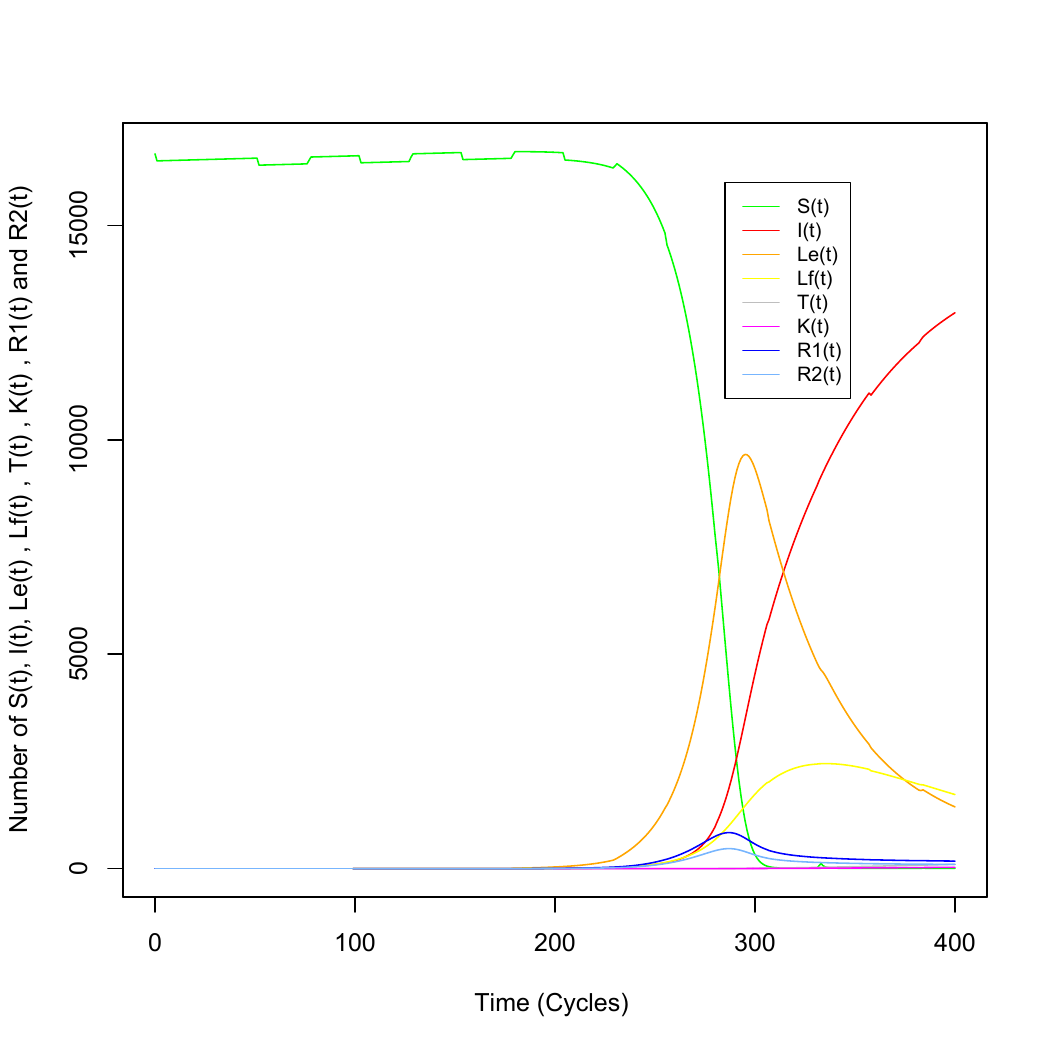} \\
		(c) & (d) \\
	\end{tabular}
	\caption{Evolution of the model with 6 cities (circles) interconnected via a hub. (a) presentation of the TB dynamics in city 1, (b) for city 2, (c) for city 3 and (d) for city 4. \label{fig:OneByOne}}
\end{figure}

The results obtained are shown in Figure~\ref{fig:AllCitya} for all the cities (global) and in Figure~\ref{fig:OneByOne} for each city separately (one by one city):

\begin{figure}[ht!]
	\begin{center}
		\includegraphics[scale=0.6]{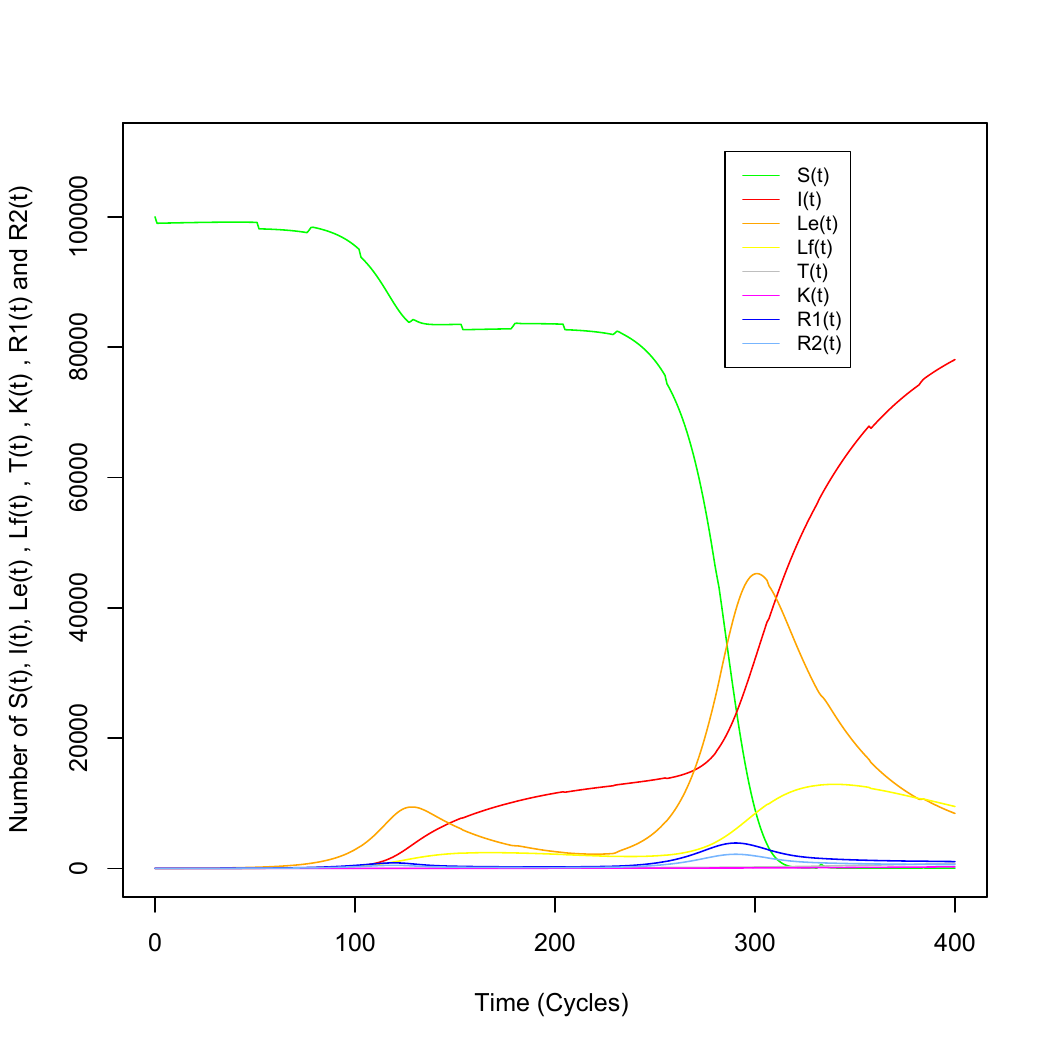}
		\caption {Evolution of the model for all cities considered globally}. \label{fig:AllCitya}
	\end{center}
\end{figure}

\subsubsection{\textit{Simulation with Basic Parameters and Consideration of a Low Mobility Rate for $I, L_e$ and $L_f$ Individuals}}

\begin{figure}[ht!]
	\centering
	\begin{tabular}{cccc} 
		\includegraphics[scale=0.2]{cycle0.png}&
		\includegraphics[scale=0.2]{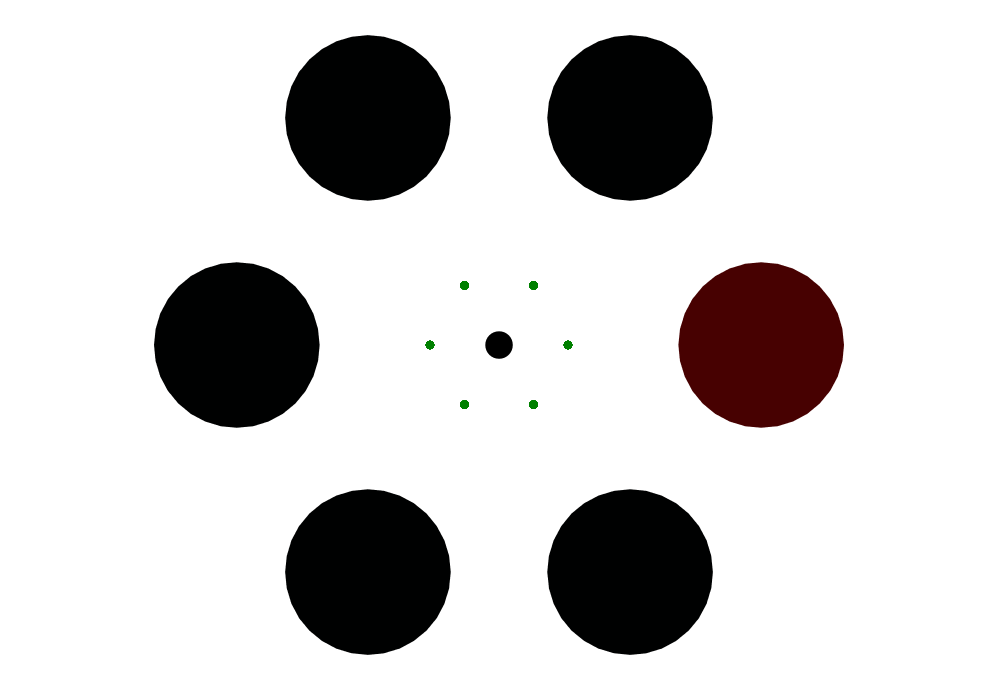} \\
		(a) & (b) \\
		\includegraphics[scale=0.2]{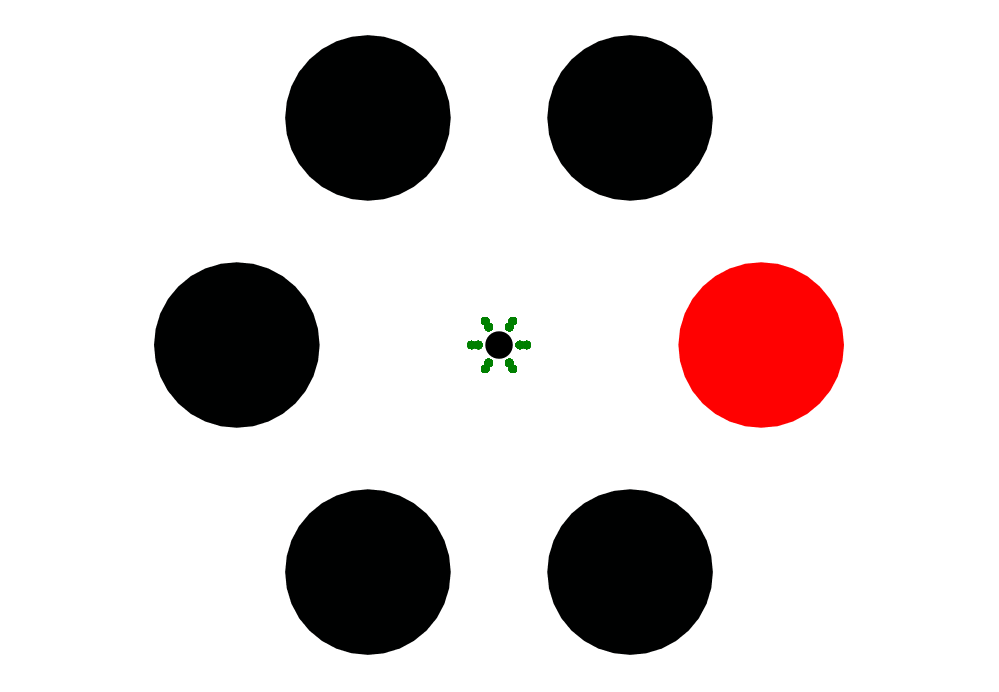}&
		\includegraphics[scale=0.2]{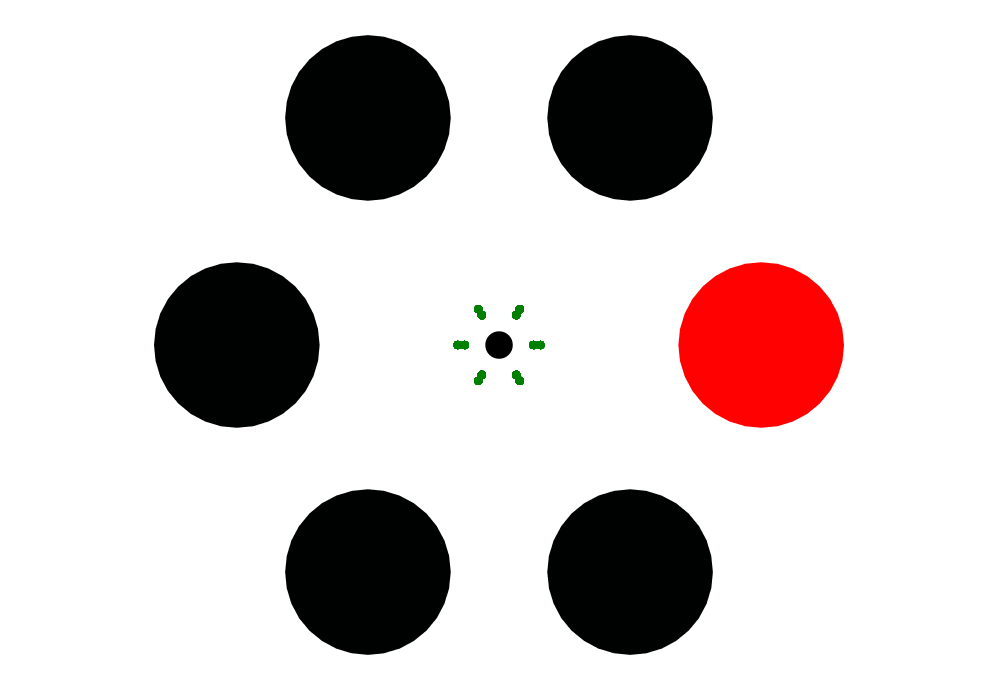} \\
		(c) & (d) \\
	\end{tabular}
	\caption{Environment with 6 cities (circles) interconnected via a hub. (a) simulation at time $t=0$, (b) simulation at time $t=79$, (c) simulation at time $t=262$, (d) simulation at time $t=328$. The concentration of the red color of the city is proportional to the level of infectivity of the city. \label{fig:AllCity}}
\end{figure}

The environment with 6 cities connected via a hub is presented in Figure~\ref{fig:AllCity}. The mobility rate considered for $I$, $L_e$ and $L_f$ is equal to $0.00001$. 

\begin{figure}[ht!]
	\begin{center}
		\includegraphics[scale=0.45]{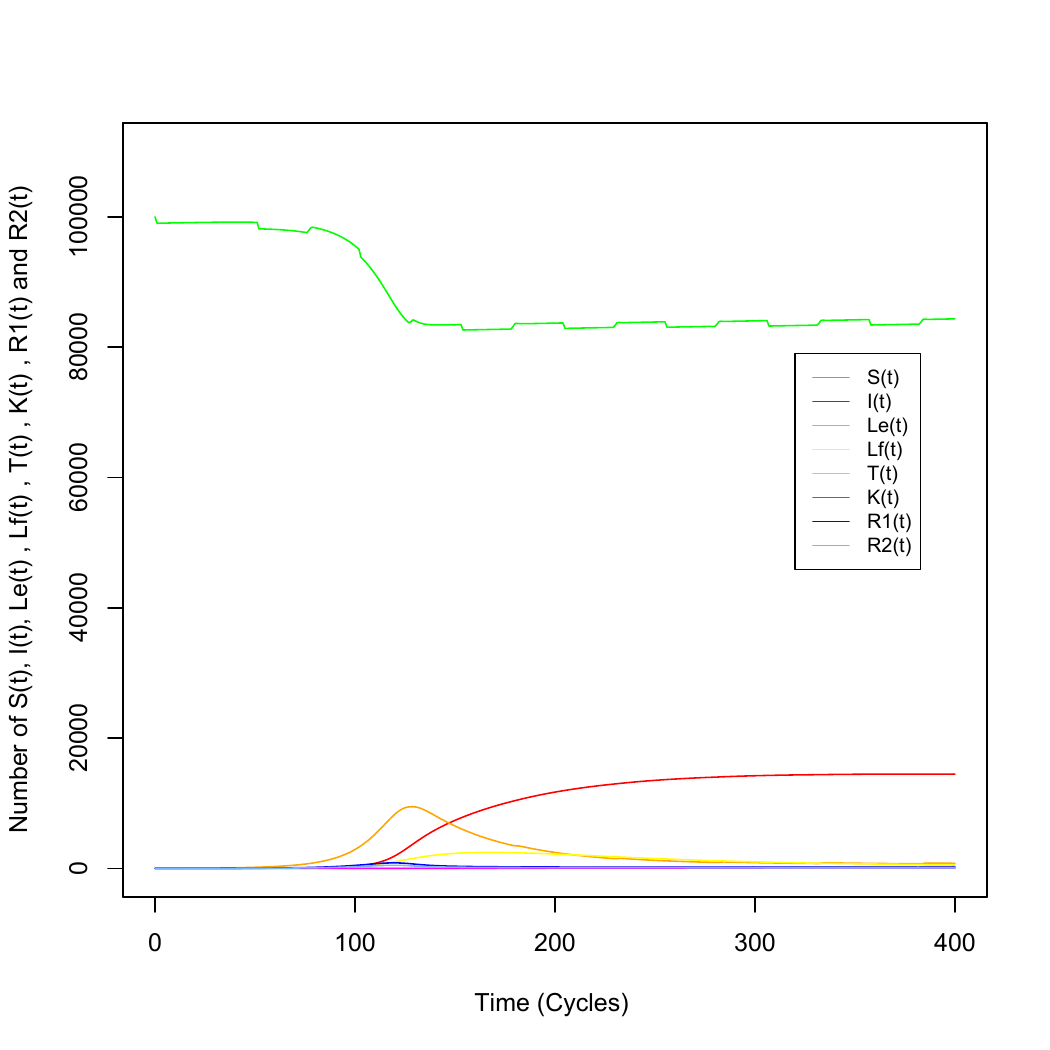}
		\caption {Evolution of the model for all cities considered globally.} \label{fig:AllCity1}
	\end{center}
\end{figure}

\begin{figure}[ht!]
	\centering
	\begin{tabular}{cccccc} 
		\includegraphics[scale=0.37]{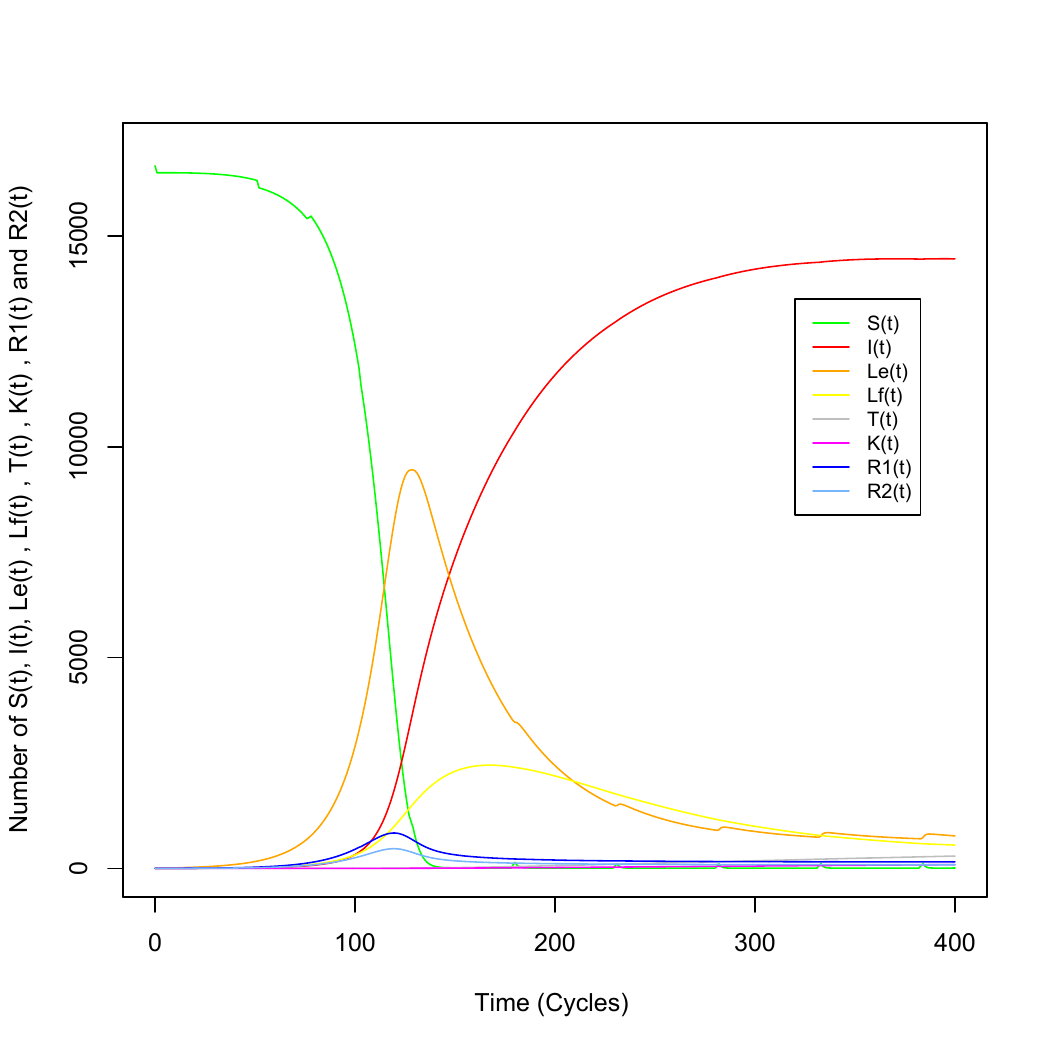}&
		\includegraphics[scale=0.37]{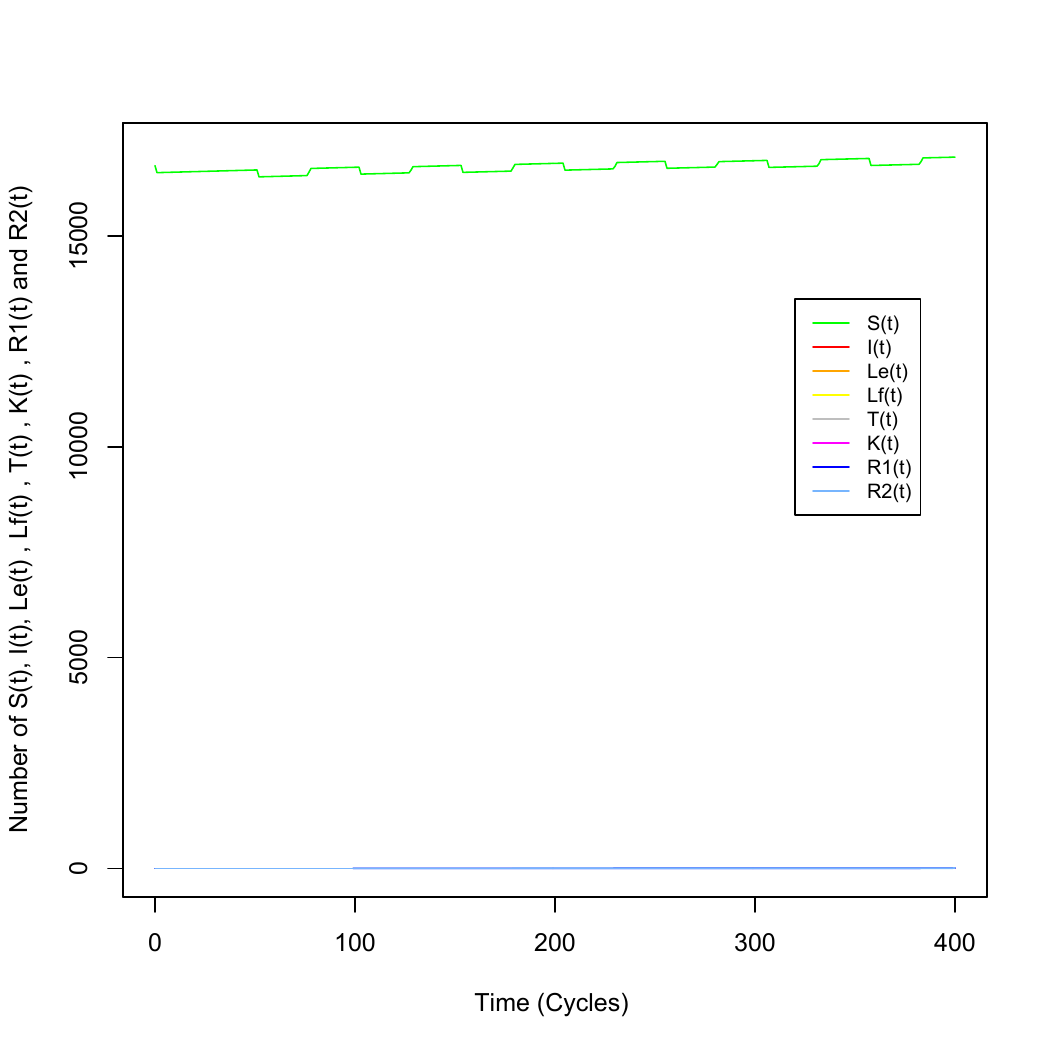} \\
		(a) & (b) \\
	\end{tabular}
	\caption{Evolution of the model with 6 cities (circles) interconnected via a hub. (a) show the dynamics for city 1 and (b) for city 2. \label{fig:OneByOne1}}
\end{figure}

The results obtained are shown in Figure~\ref{fig:AllCity1} for all the cities and in Figure~\ref{fig:OneByOne1} for each city separately.

\subsection{Runtime Evaluation \label{Runtime}}

In the modeling and simulation of infectious disease spread, combining equation-based models and agent-based models allows for the creation of a hybrid model that benefits from the advantages of both approaches. This combination notably reduces execution time and decreases resource requirements, unlike a model entirely based on agents.

In Table~\ref{Tab:Runtime}, we present the execution time of the ABM \cite{kabunga2020stochastic}, EBM \cite{kasereka2020analysis}, and Hybrid model separately with different values for the susceptible individuals $(S)$. For the ABM, the mixed population structure is considered, while for the hybrid model, the population structure is a network with three cities connected via a central hub. The simulation concludes when the number of individuals infected with tuberculosis $(I)$ reaches zero ($(I=0)$). This means that the disease is eliminated in the population.

\begin{table}[h!]
\footnotesize
\centering
\caption{Runtime for the EBM, ABM and Hybrid Model according to the number of  susceptible individuals (S) and the basic reproduction number $\mathcal R_0 = 0.6887$ for the mathematical and hybrid models.\label{Tab:Runtime}}

		\begin{tabular}{p{0.10 \textwidth} p{0.2 \textwidth} p{0.2 \textwidth} p{0.2 \textwidth}}
			\toprule
			  \textbf{S} & \textbf{ABM} & \textbf{Hybrid Model} & \textbf{EBM}\\
			\midrule
 50 & 17 cycles & 11 cycles & 8 cycles  \\
   100      & 34 cycles  & 20 cycles & 13 cycles   \\
   250     & 58 cycles  & 30 cycles & 18 cycles  \\
   500      & 82 cycles & 55 cycles & 23 cycles  \\
 750 & 87 cycles  & 65 cycles & 29 cycles   \\
   1000      & 92 cycles  & 80 cycles & 34 cycles   \\
  1250      & 103 cycles  & 90 cycles & 37 cycles   \\
  2000       & 110 cycles  & 100 cycles & 43 cycles \\     
  3000 & 115 cycles & 105 cycles & 49 cycles  \\
   4000      & 120 cycles  & 111 cycles & 54 cycles   \\
   5000     & 124 cycles  & 115 cycles & 58 cycles   \\
   6000      & 126 cycles  & 118 cycles & 70 cycles   \\
 7000 & 129 cycles  & 120 cycles & 75 cycles   \\
  8000       & 132 cycles  & 125 cycles & 88 cycles  \\
  9000      & 136 cycles  & 128 cycles & 90 cycles  \\
   $10^{5}$      & 140 cycles  & 131 cycles & 100 cycles \\
\bottomrule
		\end{tabular}
\end{table}

\begin{figure}[h!]
\centering
\includegraphics[scale=.25]{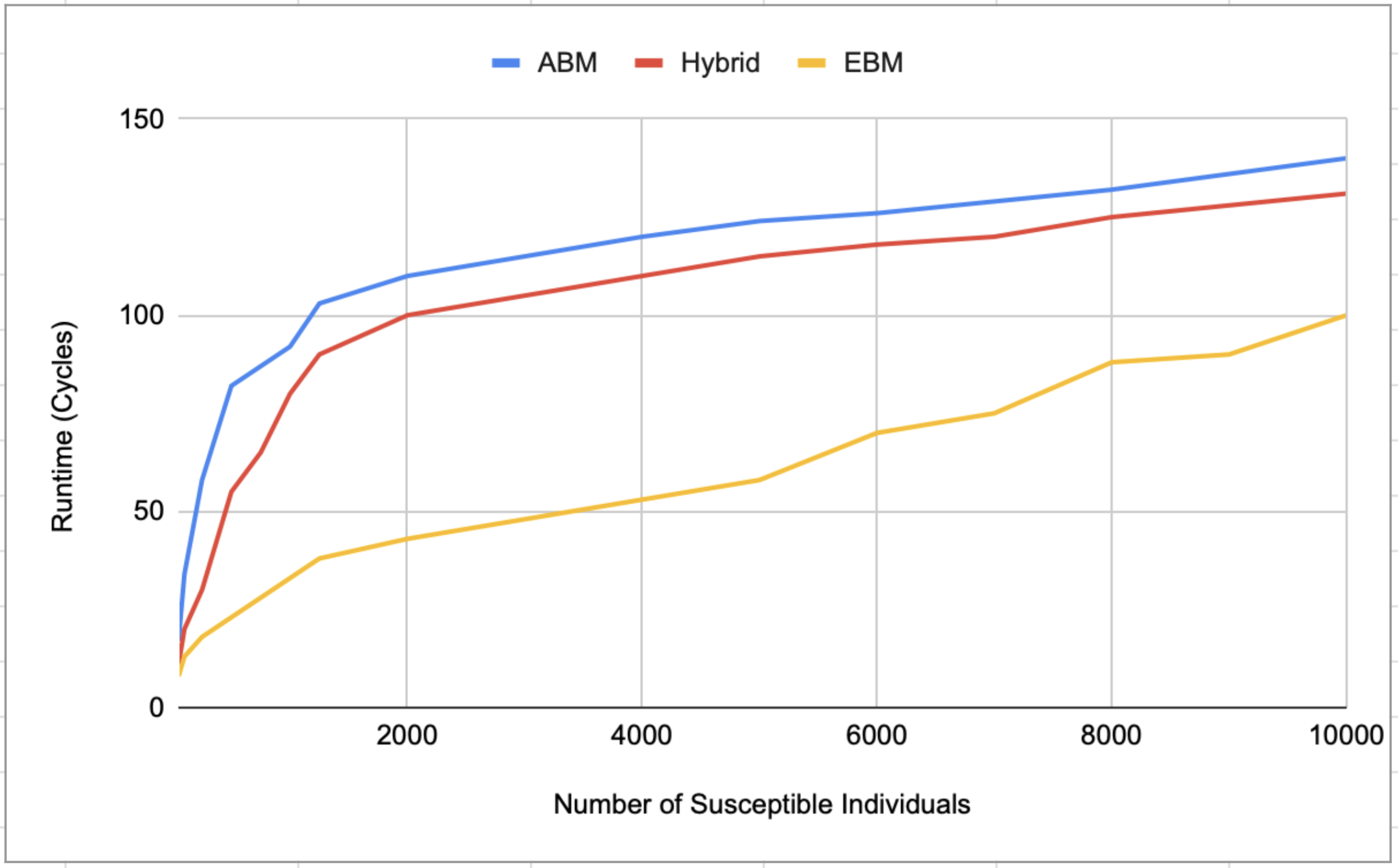}
\caption{Runtime of ABM, EBM and Hybrid model according to the number of the susceptible individuals. }
\label{Fig:RunTimeAll}
\end{figure}

Based on the results presented in Table~\ref{Tab:Runtime}, we present in Figure \ref{Fig:RunTimeAll}, the evolution of the runtime of each model according to the number of the susceptible individuals.

An overall comparison of the execution times of these models shows that the hybrid model reduces the execution time by around 17.86\% compared with the ABM model. It should be noted that the simulations were conducted on a MacBook Pro equipped with an Intel Core i7 6-core processor, clocked at 2.2 GHz. This computer features a 256 KB L2 cache per core, a 9 MB L3 cache, and Hyper-Threading technology enabled. It has 16 GB of 2400 MHz DDR4 RAM and a Radeon Pro 555X graphics card with 4 GB of memory.

\subsection{Benchmark Models}

Several reasons have confirmed the need to couple agent-based models with differential equation-based models to understand disease spread at different scales. Agent-based models make it possible to simulate complex interactions between individuals, which are difficult to capture with differential equations alone. For example, the spread of a disease may depend on a variety of individual behaviors and specific social contacts. Differential equations are effective for modeling large-scale dynamics, such as the global spread of an epidemic. In contrast, agent-based models can simulate small-scale dynamics, such as interactions within a community. It should be noted that agent-based models can be easily adapted to include new behaviors or agents, while differential equations offer a rigorous mathematical structure for analyzing global trends. On the basis of the motivations set out above and the evidence presented in the Table~\ref{Tab:compare}, it appears that coupling these two approaches has many advantages.

\begin{table}[h!]
\tiny
\caption {Brief comparison table of Equation-Based Model and and Agent-Based Model.}
\label{Tab:compare}
\begin{center}
\begin{tabular}{p{0.30 \textwidth}  p{0.30 \textwidth} p{0.3\textwidth}}
			\toprule
\textbf{Criteria} & \textbf{EBM} & \textbf{ABM} \\ 
\toprule
Nature & Synthetic: models like the SIR compartmental model simplify complex biological processes into equations.
  & Many parameters: to simulate individual entities, such as agents in an ecosystem. \\  
\midrule            
Mathematical Resolution &  Formalized: the model uses formalized mathematical approaches like differential equations to describe system dynamics & Non-formalized: the model relies on non-formalized, rule-based interactions to simulate behaviors and interactions of agents.
\\                   
\midrule
Biological Abstraction  & Far from abstract biology: the model abstracts biological processes to a high level, often simplifying them significantly. & Close to biology: the model maintains a closer representation of biological processes, simulating detailed behaviors of individual entities.   \\               
\midrule
Modularity and Incrementality & Little modular and incremental: the model is typically less modular and incremental, with fixed equations that are harder to modify. & Modular and incremental: the model is highly modular and incremental, allowing for easy addition of new behaviors and interactions.
    \\  
\midrule     
Description Level & Population level: the model describes phenomena at the population level, such as overall infection rates in a population.
 & Entity level (cell/molecule): the model describes phenomena at the entity level, such as the behavior of individual cells or molecules.   \\   
\midrule
Abstraction Level & High abstraction level, focusing on aggregated data and overall trends. & Low abstraction level, focusing on detailed, individual behaviors and interactions.  \\ 
\midrule                           
Runtime  & Low: the model generally has a low runtime, suitable for quick simulations of simple models.  & High: the model often has a high runtime due to the complexity of simulating numerous individual entities. \\            
\midrule              
System Homogeneity  & Homogeneous: the model ssumes homogeneity within the population, with uniform characteristics. & Heterogeneous: the model takes heterogeneity into account, with diverse characteristics and behaviors among agents.  \\
\midrule
Computational Requirements  & Low: the model requires minimal computational resources, suitable for simpler models.
 & High: the model requires significant computational power, especially for large-scale simulations.  \\                    
\midrule
Scalability  & High: the model can handle large populations efficiently due to its aggregated nature.   & Moderate: scalability can be challenging due to the detailed simulation of individual entities.  \\       
\midrule             
Ease of Implementation  & Moderate: the model requires a solid understanding of differential equations and mathematical modeling. & Complex: the model involves complex programming and detailed parameter settings.  \\ 
\midrule                   
Flexibility  & Low: the model is less flexible as it relies on predefined equations.
  & High: the model is highly flexible, allowing for the inclusion of diverse behaviors and interactions.  \\                       
\bottomrule
\end{tabular}
\end{center}
\end{table}

To demonstrate the robustness of our approach, including the advantages of combining the agent-based model with the equation-based model, we present in Table~\ref{Tab:benchmark} a comparative study with various compartmental, agent-based, and hybrid models found in the literature. 

In this comparative study, we highlight the objectives pursued by the different authors and highlight several aspects such as stochasticity (yes/no), multi-scale environment (yes/no), runtime (Very low/Low/High) and the limitations of each model. 

\begin{table}[h!]
\tiny
\caption{Benchmarking of the models, including the approach used, the objective of the study, the stochastic aspect of the model, the runtime and the limitations of the study.\label{Tab:benchmark}}

		\begin{tabular}{p{0.10 \textwidth} p{0.25 \textwidth} p{0.05 \textwidth} p{0.05\textwidth} p{0.07 \textwidth} p{0.30 \textwidth} p{0.05\textwidth}}
			\toprule
			 \textbf{Model} & \textbf{Purpose} & \textbf{ST} & \textbf{MS} & \textbf{RunT}  & \textbf{Limits}  &\textbf{Refs.}\\
			\midrule
 ABM & Study the dynamics of TB transmission and the role of various contact networks in the disease using an Agent-Based Model. &   Yes &  Yes &  High & The individual’s travel time from a source to a chosen destination is not clearly specified. So, it is not easy to assess the impact of the population mobility on the evolution of TB incidence in different contact networks. & \cite{kasaie2013agent} \\
\midrule
 EBM &   A mathematical compartmental model is proposed to understand the spread of TB, including groups lost to follow-up and those who have been transferred. The model's results indicate that treating latent TB infections and managing population mobility are crucial strategies that TB management team should consider to eliminate the disease in the country. &  No & No &  Very low & The mathematical model presented here offers a comprehensive overview of the spread of TB at a population level. However, it does not delve into the intricate interactions between individuals, which can significantly influence the dynamics of disease transmission. By focusing on a global perspective, the model overlooks the heterogeneity in individual behaviors, contact patterns, and the impact of localized interventions. The model may not fully capture the nuances required for targeted interventions and personalized treatment plans. & \cite{kasereka2020analysis} \\
\midrule
 EBM \& ABM &  A hybrid model is proposed for managing the spread of infectious diseases. The proposed model allows geographic areas to switch between the equation-based model and the agent-based model depending on the number of agents infected in the environment. The model was tested at the town level and then at the county level to compare the time saved compared to a fully agent-based model. &   Yes & Yes &  Low & As the switch between ABM and EBM depends on the number of infected agents, if the number of these agents is very low, the transition will not take place at that time. In this case, the hybrid model will not capture the heterogeneous behavior of these agents, which could influence and contribute to the dynamics of the disease under consideration. In addition, since the compartmental model considered is of the simplistic SEIR type, it is difficult to confirm whether the amount of fidelity will not be lost. & \cite{hunter2020} \\
\midrule
  EBM \& ABM &   The proposed model integrates two modeling paradigms, ABM and EBM, to create a hybrid approach that optimizes computational efficiency under varying conditions. Initially, the model operates as agent-based, but it transitions to an equation-based approach once the number of infected individuals becomes sufficiently large to justify a population-averaged method. &  Yes &   Yes & Low & As the model starts as agent-based and switches to equation-based when the number of infected individuals is large, the run time should remain high compared to other hybrid models if the ABM is complex. Additionally, the presented example used a simplified Agent-Based Model and does not account for behavior change during an epidemic. &  \cite{4419767}\\
\midrule
  EBM \& ABM &  In this model we propose a hybrid model that combines an EBM with an ABM at different scales. This approach allows us to capture both the broad trends and detailed interactions within the population. By simulating individual movements and behaviors, the model provide insights into how mobility influences TB transmission in a multi-scale environment. &    Yes &  Yes &  Low & The mathematical model employs eight compartments to analyze TB dynamics at the population level, which adds to its complexity. Additionally, the proposed experimental model does not utilize real-world data. &  Our model\\
\bottomrule
		\end{tabular}
	\noindent{\tiny{ST : Stochasticity, MS: Multi-scale and RunT : Runtime}}
\end{table}


\section{Discussion of the Results and Recommendations \label{Discussion}} 

\subsection{General Discussion}

The hybrid model implemented and simulated shows that individual mobility has an impact on the spread of tuberculosis in the population. Indeed, Figs.~\ref{fig:AllCitya} and~\ref{fig:OneByOne} illustrate that when there is a large number of people moving from city to city (mobility rate equal to 0.01 of the population according to its status), the disease spreads to all cities. This is justified by the fact that infected persons moving from their home city to another city contribute to the epidemiology of the destination city (randomly selected). The persistence of tuberculosis in poor countries can be partially justified by the rural migration. For example, in the Democratic Republic of the Congo, poor people living in the provinces move to the capital (Kinshasa) to seek a better life. Kinshasa being already one of the cities most affected by TB~\cite{bisuta2018tendances}, the arrival of new cases from other cities is one of the factors contributing to the persistence of tuberculosis in this city.

By reducing the mobility rate to 0.00001 for people with TB infection and people with latent TB, simulations show that the disease does not spread in all cities as shown in Figs.~\ref{fig:AllCity1} and~\ref{fig:OneByOne1}. This means that restricting the mobility (between cities) of infectious and latent people is one of the measures to be taken to control the spread of tuberculosis across a country.

Based on the results obtained, it seems that the mobility of individuals from city to city has a significant impact on the spread of tuberculosis in the population. Indeed, when people move from one city to another, they can carry tuberculosis bacteria with them, potentially spreading the infection to new areas. This is of particular concern in regions with high TB prevalence, or where health systems are not well coordinated. Short-term mobility between different areas can lead to an increase in TB prevalence, as individuals can move from a high-risk to a low-risk area, potentially introducing new infections. In addition, mobility can disrupt continuity of care for those undergoing TB treatment, leading to poorer health outcomes. TB control efforts need to take account of these mobility patterns in order to effectively manage and reduce the spread of the disease. 

Improving the living conditions of the population in the provinces and decentralizing TB care center can be a palliative solution to the rural migration. With this strategy, people can simply stay in their home cities because the living and care conditions are the same as in other cities.

\subsection{Benchmarking Discussion}

Simulation results show that our approach is a compromise that benefits from the advantages of both ABM and EBM. Table~\ref{Tab:Runtime} and Figure \ref{Fig:RunTimeAll} show that the proposed hybrid model reduces the execution time compared with the ABM model. 

Based on the comparative study presented in Table~\ref{Tab:benchmark}, it appears that the combination of these two types of modeling offers a more complete and nuanced understanding of the spread of tuberculosis in a highly mobile population and in a multi-scale environment. Comparing our approach with compartmental models, it seems that mathematical modeling offers only a global view of disease spread (macroscopic), without highlighting the interactions between individuals infected or not with tuberculosis (microscopic). However, we have found that the execution time of these models remains low. It should be noted that agent-based modeling has proven its performance in modeling interactions between individuals, taking into account the multi-scale environment and also stochasticity. On the other hand, these models require several parameters and their execution time is high compared with a hybrid model. 

This innovative approach is a powerful tool that enables managers to develop effective control strategies in multi-scale environments.

\subsection{Recommendation for Managing Population Mobility in Low and Middle Income Countries with High TB Prevalence \label{Recommendations}}

According to the results of our study, managing the mobility of people in a country with a high prevalence of tuberculosis is an important task for public health decision-makers. As rural exodus is a specific form of mobility, it is important to understand its causes, as it promotes the spread of the disease.

Since many factors associated with poverty increase the risk of tuberculosis propagation, in developing countries, poverty and lack of economic opportunity drive people to seek better living conditions in cities. This leads to high human mobility and facilitate the spread of tuberculosis on a large scale. Limited access to basic services is another cause of high population mobility from villages to cities. Indeed, the lack of essential infrastructures such as drinking water, electricity and healthcare encourages rural populations to migrate to the cities. It has also been observed that armed conflicts and insecurity in certain rural areas force inhabitants to flee to safer areas, bringing disease germs with them to the developed cities and thus contributing to the prevalence and incidence of tuberculosis in these cities.   

To effectively manage the mobility of people caused by the above-mentioned problems, governments can consider a number of solutions, including:

\begin{itemize}
\item Investing in basic infrastructure (roads, schools, hospitals) to enhance the quality of life in rural areas;

\item Promoting local initiatives and economic development projects to create jobs in rural areas;

\item Introduce sustainable farming techniques and climate-resistant crops to boost agricultural productivity; 

\item Implement measures to guarantee the security of rural populations and reduce conflict.
\end{itemize}

The fight against poverty is therefore essential to reduce the spread of tuberculosis and enhance health outcomes in affected communities. It is therefore essential to enhance access to healthcare for migrant populations, and to implement tuberculosis awareness and prevention programs.


\section{Concluding Remarks \label{Conclusion}}

In this paper, we presented a model of tuberculosis spread and simulated it on an environment composed of several cities and having a central point as a mandatory passage for the choice of the destination city for travellers. The disease dynamics at the level of each city was represented by an 8-compartment model that manages the spread of TB by solving an ordinary differential equation using the fourth-order Runge-Kutta method. To ensure the mobility of people from city to city, individuals and cities have been considered as intelligent agents. At each time step, cities generated individuals who had to travel from one city to another randomly selected according to a mobility rate. The implemented model is called hybrid because it couples the differential equations applied in each city to the intelligent agents that act on the individual by considering his heterogeneous behaviors such as mobility, random choice of destination (decision making), change of status (interaction with other agents), etc. This model benefits from the advantages of these two modeling-simulation approaches for complex systems. The model implemented here has the advantage of not requiring a large number of parameters and a high capacity in terms of machine power. In addition to the advantage of being synthetic, this hybrid model benefits from the analytical character of mathematical models. Agent-based modeling enriches this model by imposing a stochastic character on it through random facts introduced into the random choice of the destination city for individuals but also into the generation of individuals in different cities at a randomly chosen rate. 

The consideration of the individual as an entity in its own right and the management of his or her behaviour gives the microscopic aspect to the model set up and brings it as close as possible to reality. The mathematical management of the spread of the disease in cities gives a macroscopic aspect to the model. This model therefore draws its richness from this dynamic at two different scales, which gives the emergence of the model at the global level. As a result, it seems obvious that the coupling of mathematical models to agent-based models will be applied when the dynamics of the complex system under consideration is at different scales. The implemented hybrid model does not require enough computing power compared to the general agent-based model which requires high performance simulation hardware. The results obtained show that coupling the EBM and ABM approaches is an interesting compromise, as confirmed by our latest work described in ~\cite{kasereka2023equation}.

As part of the outlook, we intend to adjust the proposed model and proceed with several applications:

\begin{itemize}

\item Use real tuberculosis data to compare the results obtained with those of simulated data. This should help us validate the proposed model;

\item Replace the ordinary differential equations used in the hybrid model with stochastic equations, which should make it possible to introduce randomness at all levels of population and disease dynamics, making the proposed model even more realistic;   

\item Integrate the TB vaccination at the microscopic level of this model. This should make it possible to see its impact at the global level of the model.
\end{itemize}

Finally, we believe that the hybrid approach presented in this paper is versatile and can be applied to understand the dynamics of other diseases without requiring significant modifications.    

\section*{Acknowledgements}
The author would like to thank Professors Emile-Franc Doungmo Goufo, Ho Tuong Vinh and Kyandoghere Kyamakya for their valuable discussions during the preparation of this paper. He expresses his deep gratitude to the referees for their valuable suggestions regarding the revision and improvement of the manuscript. The author also extends his thanks to Professor Ruffin-Beno\^it M. Ngoie for proofreading the manuscript.

\section*{Funding}
This research received no specific grant from any funding agency in the public, commercial, or not-for-profit sectors.

\section*{Competing Interests}
The author declares that he has no known competing financial interests or personal relationships that could have appeared to influence the work reported in this paper.

\section*{CRediT Author Statement}
The author confirms sole responsibility for the following: study conception and design, methodology, software development, numerical simulations, analysis and interpretation of results, and manuscript preparation.

\section*{Code Availability}
The GAML (GAma Modeling Language) source codes developed for the proposed hybrid model can be found at \url{https://github.com/sedjokas/TB_Hybrid_Model} (accessed on 8 october 2024).

%
%
%
\bibliographystyle{elsarticle-harv}
\bibliography{elsarticle-template-num}
%





\end{document}